\documentclass[slac_one]{revtex4}
\usepackage{graphicx}
\usepackage{fancyhdr}
\def\H4mu{\ensuremath{\mathrm{H}\to\mathrm{ZZ^{(\ast)}}\to4\mu}}

\begin{document}

\title{Measurement of the Charge Ratio of Cosmic Muons using CMS Data} 
\author{M. Aldaya\footnote{Currently at DESY, Notkestr. 85, 22607, Hamburg, Germany}, P. Garcia-Abia (for the CMS collaboration)}
\affiliation{CIEMAT, Avda. Complutense 22, E-28040, Madrid, Spain}

\begin{abstract}
We have performed the measurement of the cosmic ray muon charge ratio, as a function of the muon momentum, using data collected by the CMS experiment, exploiting the capabilities of the muon barrel drift tube (DT) chambers. The cosmic muon charge ratio is defined as the ratio of the number of positive- to negative-charge muons. Cosmic ray muons result from the interaction of high-energy cosmic-ray particles (mainly protons and nuclei), entering the upper layers of the atmosphere, with air nuclei. Since these collisions favour positive meson production, there is an asymmetry in the charge composition and more positive muons are expected. 

The data samples were collected at the \textit{Magnet Test and Cosmic Challenge} (MTCC). While the MTCC itself was a crucial milestone in the CMS detector construction, not having physics studies among its primary goals, it provided the first opportunity to obtain physics results and test the full analysis chain using real data in CMS before the LHC startup, together with a complementary check of the detector performance.
\end{abstract}

\maketitle

\thispagestyle{fancy}

\section{EXPERIMENTAL SETUP AND DATA SAMPLES}

\begin{figure}[ht]
  \begin{center}
    \resizebox{8.0cm}{!}{\includegraphics{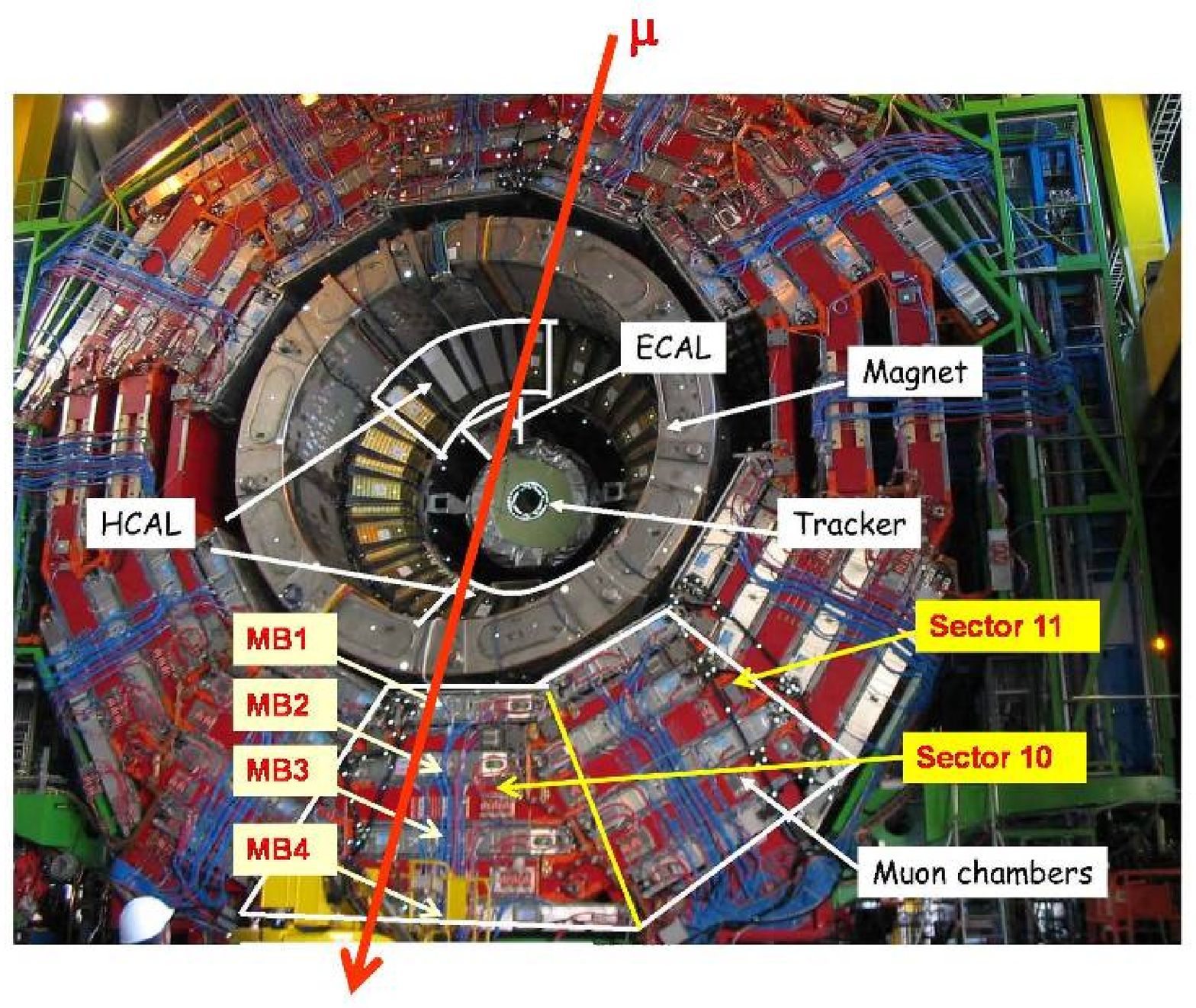}} \hfil
    \resizebox{7.0cm}{!}{\includegraphics{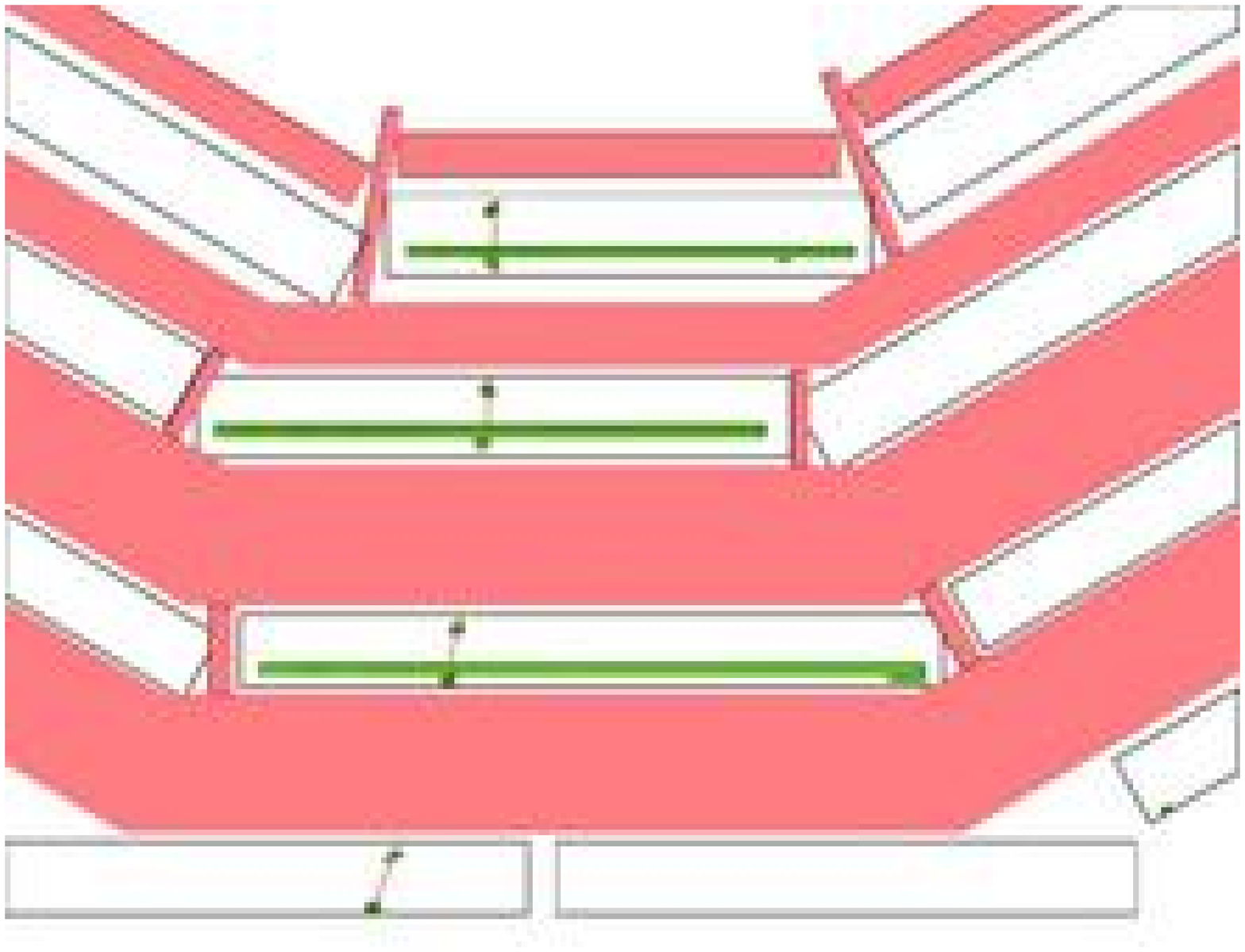}} 
    \caption{(Left)~Experimental setup of the MTCC. (Right)~A cosmic muon observed in DT Sector 10 of one of the wheels (transverse, r-$\phi$, view). The dots and horizontal lines indicate the hits in the muon system, the thin lines showing the reconstructed muon track.} 
    \label{setup-f1}
  \end{center}
\end{figure}

The MTCC~\cite{mtcc} took place in the second half of 2006. It was a first and successful major performance test of a small fraction of the CMS detector~\cite{cms}, including the magnet, using cosmic-ray muons as a particle beam, before it was lowered to the experimental cavern. The detector elements participating in the MTCC are shown in Figure~\ref{setup-f1}~(left). A cosmic muon detected in CMS during the MTCC is displayed in Figure~\ref{setup-f1}~(right). 

The analysis is performed with the DTs alone (Sector 10 of wheels YB+1, YB+2). The runs used in the analysis (5 in total) were taken at B $\sim$ 4 T, all with similar DT trigger conditions (at least MB2+MB3 stations). A control sample taken at 0 T is used to make detailed checks of detector performance. Simulated samples (B = 0 T and B = 4 T)~\cite{mc} are used for cross-checks and minor corrections.

The events are reconstructed using the standard reconstruction software of the CMS experiment. 
The so-called \textit{Standalone muon reconstruction} was adapted for dealing
with the specific spatial configurations of cosmic muons~\cite{cliu}.
The reconstruction procedure uses special DT calibration constants~\cite{cali-anna}, in order to account for the random arrival time of cosmic muons. Alignment corrections~\cite{parbol},
calculated from survey data~\cite{survey}, are applied to the measured hits before the track reconstruction.


\section{ANALYSIS AND EVENT SELECTION}

The analysis strategy is selecting a sample of high quality muons, obtaining the momentum spectra as a function of the muon charge and calculating the charge ratio as a function of the momentum. 

\begin{figure}[ht]
  \begin{center}
    \resizebox{5.5cm}{!}{\includegraphics{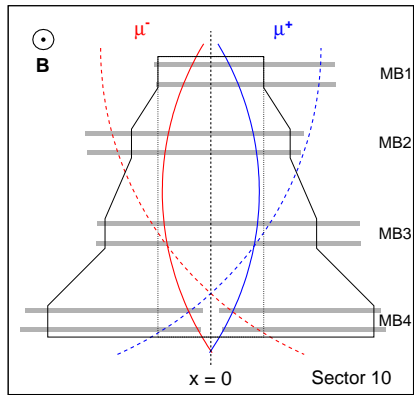}} 
    \caption{Definition of the left-right symmetric fiducial geometry (solid polynomial line) of the analysis. The dashed lines depict two muon tracks with the same momentum crossing Sector 10, the negative one satisfying the MB2+MB3 trigger condition and the positive one failing it. The solid curves represent two muons with the same $p$ in the fiducial geometry, both of them passing the selection criteria.} 
   \label{geometry-f2}
  \end{center}
\end{figure}

In order to select well-reconstructed muons and avoid bias in the relative acceptance for $\mu^+$ and $\mu^-$, caused by trigger configuration (MB2+MB3), momentum resolution and asymmetric geometrical acceptance due to magnetic field, muons are required to leave hits in at least 3 stations (triplets/quadruplets), and to be fully contained in a left-right symmetric region in Sector 10 of wheels YB+1 and YB+2. Figure~\ref{geometry-f2} helps understanding the influence of the geometric acceptance
in the relative efficiency for $\mu^+$ and $\mu^-$ with the same momentum, entering the detector in symmetric
points, and with the same angle, with respect to the vertical axis.

\begin{figure}[ht]
  \begin{center}
    \resizebox{7.0cm}{!}{\includegraphics{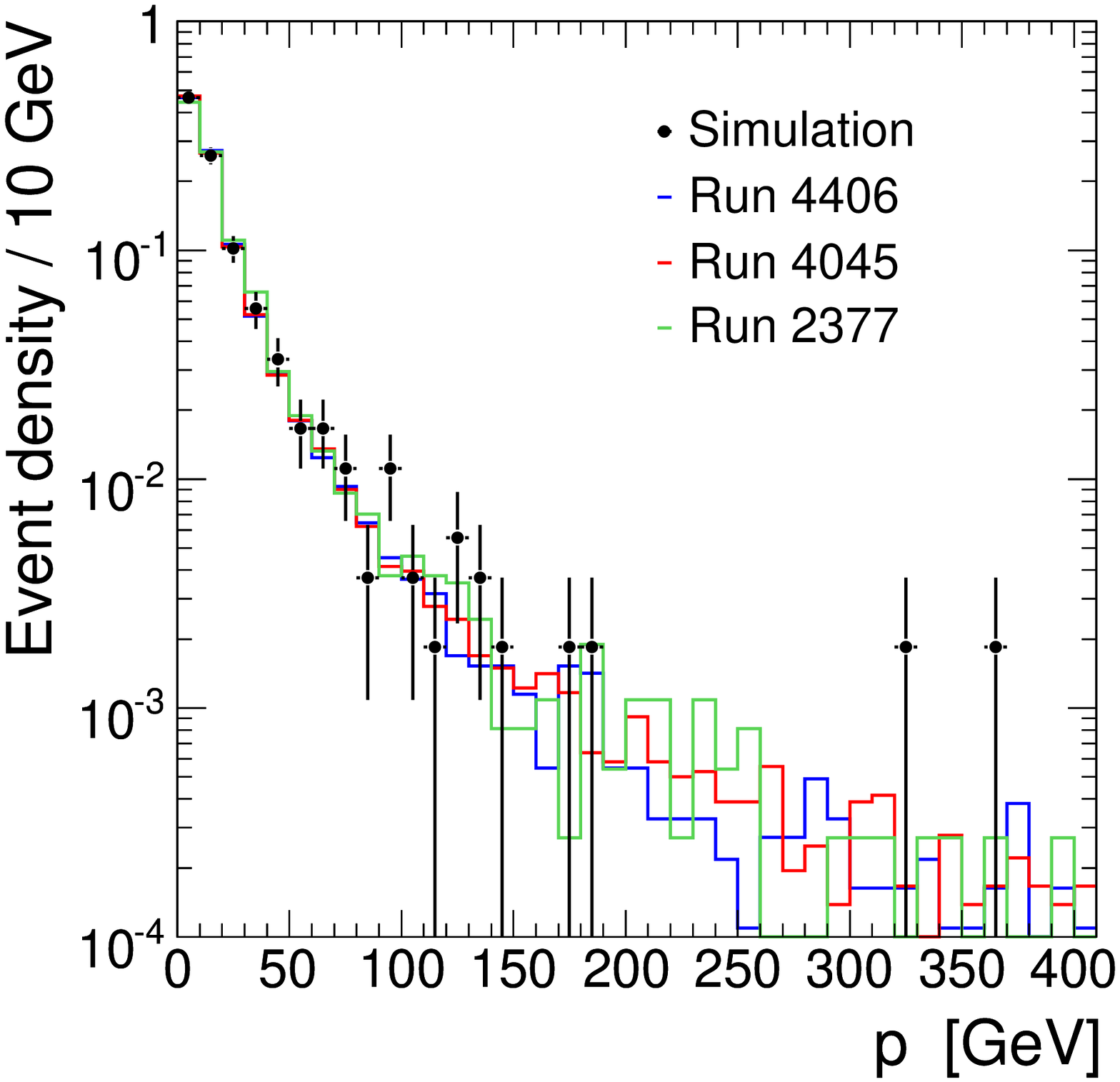}}
    \resizebox{7.0cm}{!}{\includegraphics{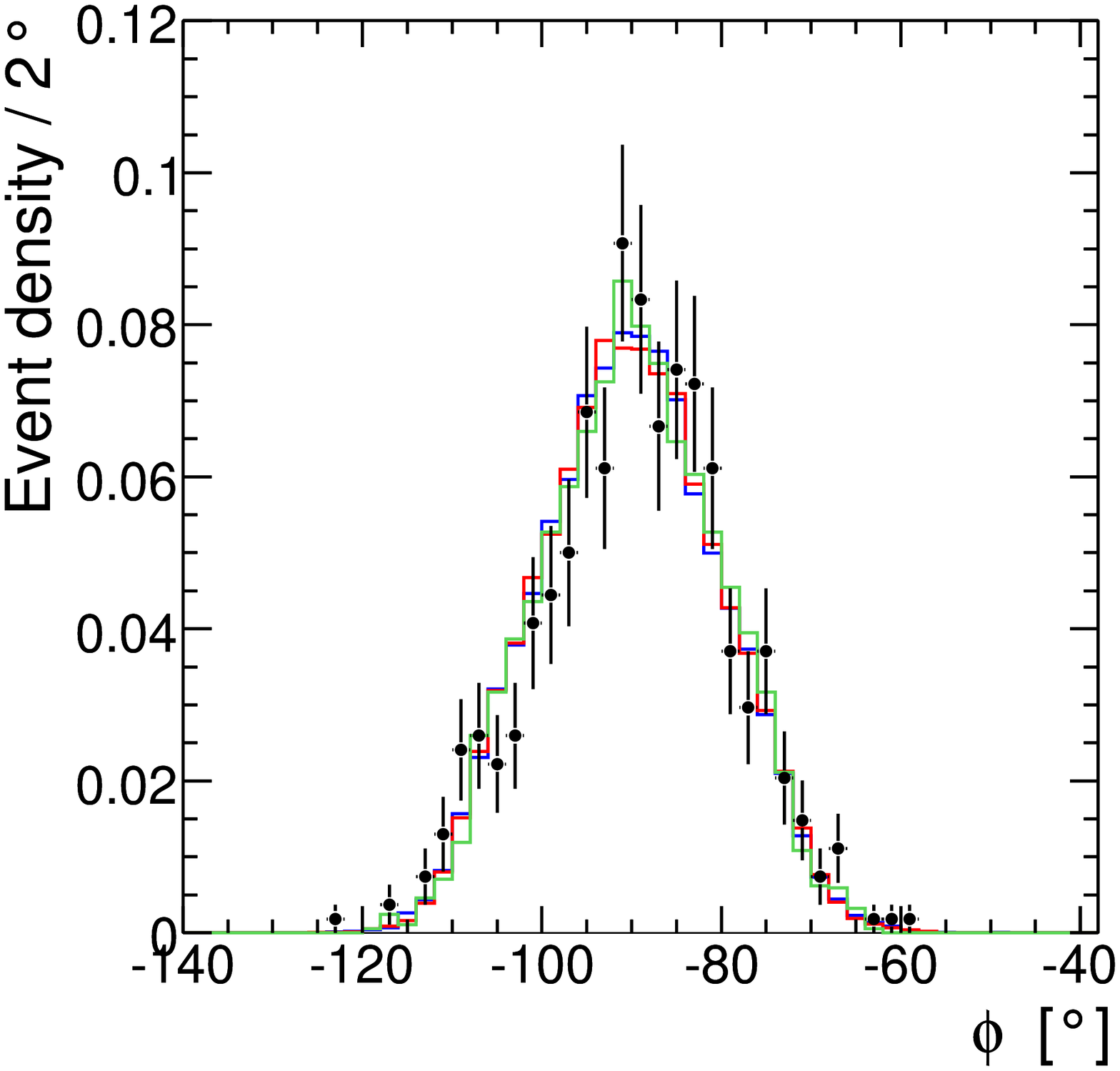}}
    \caption{Normalized distributions of the momentum of the muon tracks~(left), their azimuthal angle~(right), measured at the MB1 station, for quadruplet muons in the selected sample. The solid lines correspond to data from
             different runs and dots with error bars to the simulated events.
            }
    \label{selection_plots-f3}
  \end{center}
\end{figure}

The distributions of the momentum of the muon tracks and their azimuthal angle are shown in Figure~\ref{selection_plots-f3} for quadruplet muons in the selected sample. Good agreement between different runs and simulation is observed. The global CMS coordinate system is used, in which the $z$ axis goes along the beam-line, that is, $\phi = -90^\circ$ and $\theta = 90^\circ$ for vertical muons. 

\section{RESULTS}

The charge ratio is computed as the ratio of the $p$ spectrum of $\mu^+$ to that of $\mu^-$, as a function of the measured muon momentum ($p$), for all runs and their combination. A correction due to charge confusion and the estimation of systematic uncertainties are performed. 

\begin{figure}[ht]
  \begin{center}
    \resizebox{7.0cm}{!}{\includegraphics{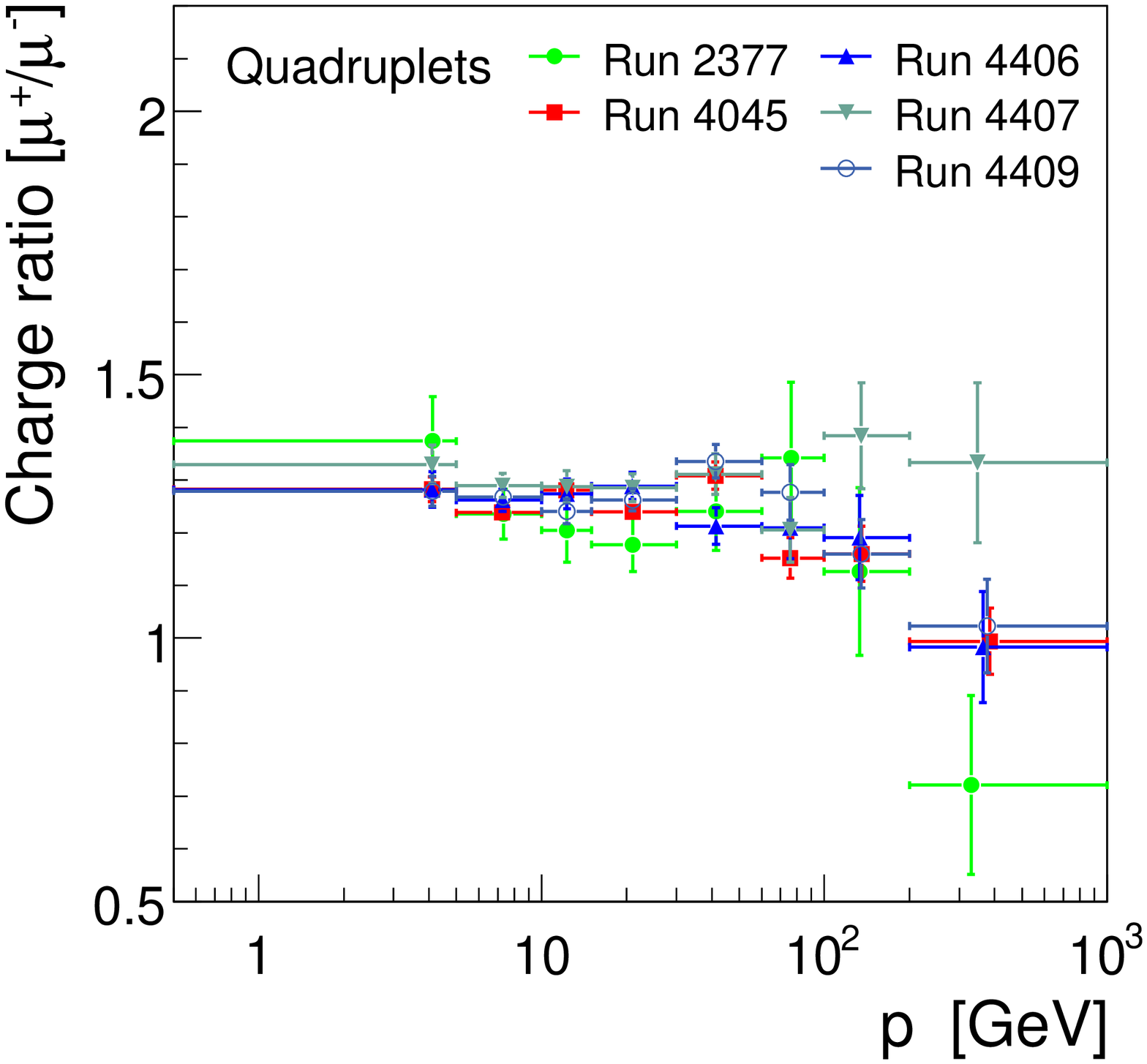}}
    \resizebox{7.0cm}{!}{\includegraphics{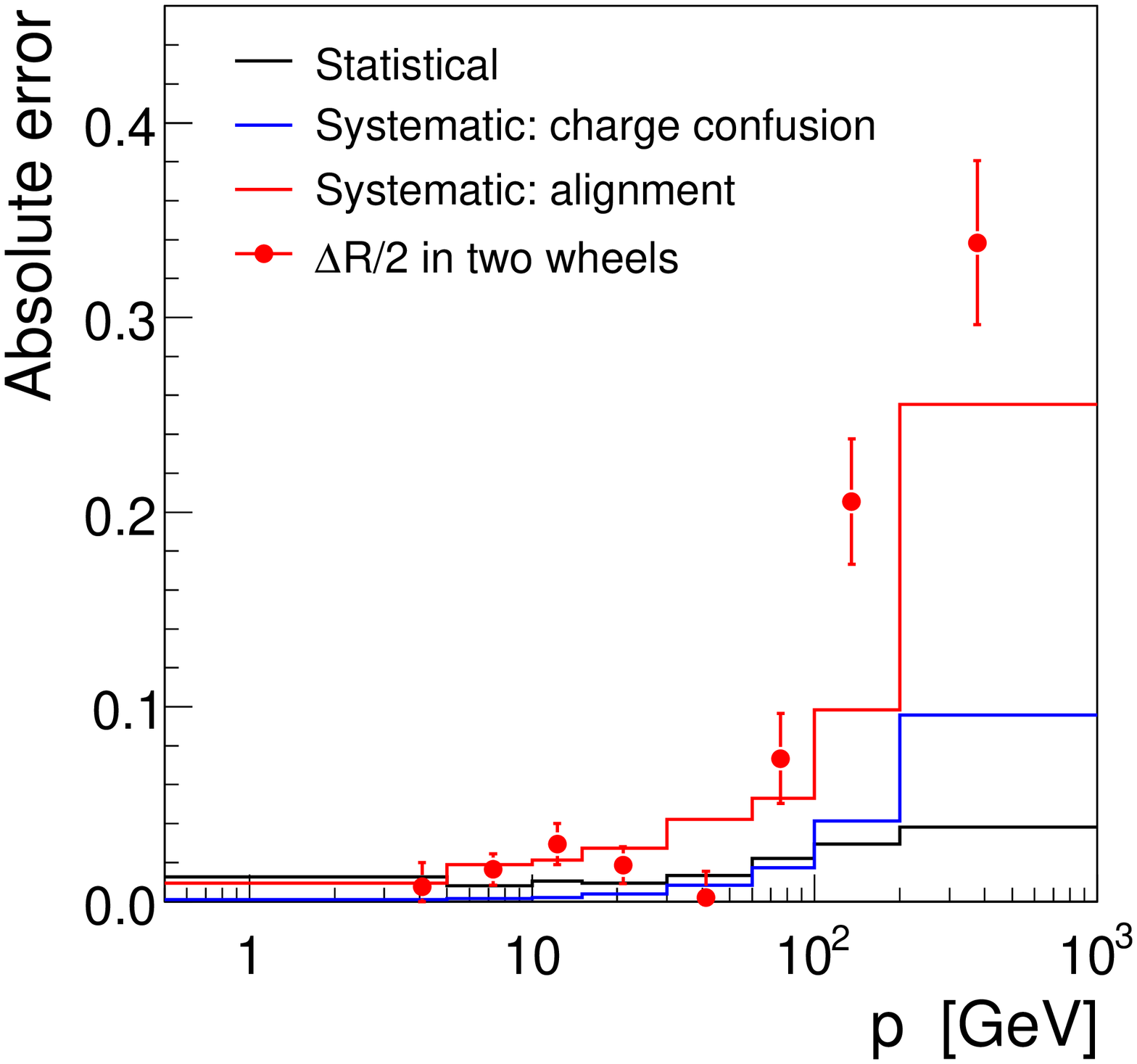}}
    \caption{(Left) Measured muon charge ratio, as function of $p$, for the individual runs for quadruplets. Only statistical errors shown. (Right) Individual contributions to the total systematic uncertainty, as function of $p$. For comparison, the half-difference of the charge ratio
             measured in the two wheels and the statistical error are also shown.}
    \label{charge_ratio_all-f4}
  \end{center}
\end{figure}

The measured muon charge ratio and its statistical uncertainty is displayed in Figure~\ref{charge_ratio_all-f4}~(left), for quadruplets, for the individual runs analyzed. 
The deviations observed among different runs are consistent with statistical fluctuations.

The individual systematic uncertainties on the charge ratio measurement are summarized in Figure~\ref{charge_ratio_all-f4}~(right), as a function of the measured muon momentum, together with the statistical error.

\begin{figure}[ht]
  \begin{center}
    \resizebox{8.0cm}{!}{\includegraphics{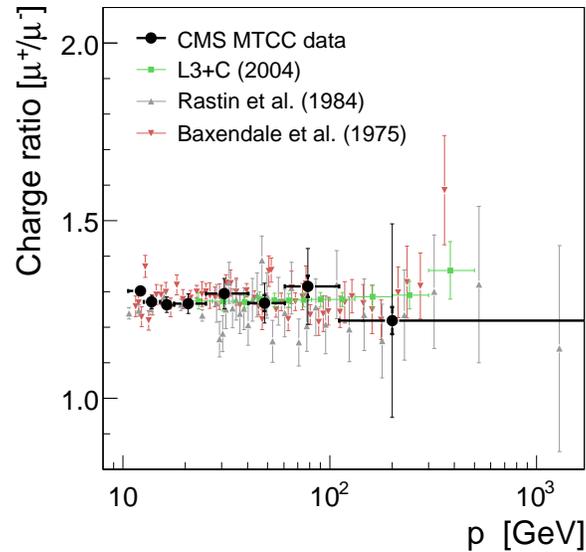}}
    \caption{Muon charge ratio measured by CMS (large dots) with statistical (thick bars) and systematic errors~(thin bars), together with results from other experiments (other markers).}
    \label{results-f5}
  \end{center}
\end{figure}

The measurement of the charge ratio using CMS data is depicted in Figure~\ref{results-f5}
as function of the measured muon momentum, together with the results from other experiments~\cite{hebbeker}. Given the experimental accuracy, the charge ratio is consistent with being independent of the muon momentum in the range of study. Under this assumption, the mean value of the charge ratio, integrated over the muon momentum, is
$ \langle R_{CMS} \rangle = 1.282 \, \pm \, 0.004 \, \mathrm{(stat.)} \, \pm \, 0.007 \, \mathrm{(syst.)} $. Good agreement with previous measurements within the experimental accuracy is observed.

This analysis is described in detail in~\cite{cmsnote}.



\begin{thebibliography}{99} 


\bibitem{mtcc}
The CMS Collaboration, ``The CMS Magnet Test and Cosmic Challenge (MTCC Phase I and II) Operational Experience and Lessons Learnt'', CMS-NOTE 2007/005.

\bibitem{cms}
The CMS Collaboration, ``Technical Proposal'', CERN/LHCC 94-38, LHCC/P1 (1994).

\bibitem{mc}
P. Biallass, T. Hebbeker, K. Hoepfner, ``Simulation of Cosmic Muons and Comparison with Data from the Cosmic Challenge using Drift Tube Chambers'', CMS-NOTE 2007/024.

\bibitem{cliu} 
C.~Liu, N.~Neumeister, ``Reconstruction of Cosmic and Beam-Halo Muons'', CMS-NOTE 2008/001.

\bibitem{cali-anna}
M.~Benettoni \textit{et al.}, ``CMS DT Chambers: Optimized Measurement of Cosmic Rays Crossing Time in absence of Magnetic Field'', CMS-NOTE 2008/017.

\bibitem{parbol}
A.~Calder\'on \textit{et al.}, ``Muon System alignment with tracks'', CMS-NOTE 2006/016.

\bibitem{survey}
J.F.~Fuchs, R.~Goudard and J.D.~Maillefaud, ``CMS-SUMMARY. YBs and YEs Position w.r.t. YB0 in SX5'', CMS-SG-UR-0490 (2006).

\bibitem{hebbeker}
T.~Hebbeker, C.~Timmermans, Astropart. Phys. 18 (2002) 107-127; \\
                       J.M.~Baxendale, C.J.~Hume, M.G.~Thompson, J. Phys. G 1 (1975) 781-788; \\
                       B.C.~Rastin, J. Phys. G 10 (1984) 1629-1638.

\bibitem{cmsnote}
M.~Aldaya, P.~Garcia-Abia, ``Measurement of the charge ratio of cosmic muons using CMS data'', CMS-NOTE 2008/016; \\
M.~Aldaya, ``Measurement of the Cosmic Muon Charge Ratio Using CMS Data and Discovery Potential of the Standard Model Higgs Boson in the $\H4mu$ Channel'', PhD. Thesis CMS-TS-2008/013. 

\end{thebibliography}
\end{document}